\documentstyle[prb,aps,epsf]{revtex}
\twocolumn
\tighten
\begin{document}
\newcommand{\lbf}[1]{\boldmath{\mbox{$#1$}}}
\draft
\twocolumn[
\hsize\textwidth\columnwidth\hsize\csname@twocolumnfalse\endcsname
\preprint{}
\title{Essential Role of the Cooperative Lattice 
Distortion in the Charge, Orbital and Spin Ordering in doped Manganites  
} 
\author{R. Y. Gu and C. S. Ting}
\address{
Texas Center for Superconductivity and Department of Physics, 
University of Houston, Houston, Texas 77204\\
}
\date{\today}
\maketitle

\begin{abstract}
The role of lattice distortion in the charge, orbital and 
spin ordering in half doped manganites has been investigated. 
For fixed magnetic ordering, we show that the cooperative 
lattice distortion stabilize the experimentally observed ordering even when
the strong on-site electronic correlation is taken into account.
Furthermore, without invoking the magnetic interactions, 
the cooperative lattice distortion alone may lead to the 
correct charge and orbital ordering including the charge stacking effect,
and the magnetic ordering can be the consequence of such a charge 
and orbital ordering.  We propose that the cooperative nature of the 
lattice distortion is essential to understand the complicated charge, 
orbital and spin ordering observed in doped manganites.

\end{abstract}
]

\narrowtext

\begin{figure}[h]
\hspace{2cm}
\centerline{\epsfxsize=3.8in \epsfbox{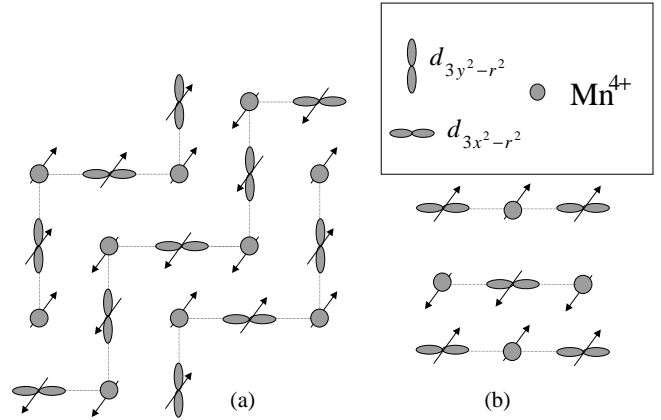}}
\caption[*]{                                                  
View of the (a) CE and (b) C phase in the x-y plane. 
The arrows refers to the spin. Along the z-direction neighboring 
sites have the same charge and orbital states but opposite spins.
}
\label{view}
\end{figure}

The unusual charge, orbital and spin ordering (COSO) in 
manganites
have recently attracted much attention 
\cite{tomioka,sternlieb,murakami,diaz,radaelli,tomioka2,kawano,solovyev,brink,yunoki,mahadevan}.
In some of these materials such as 
Pr$_{1/2}$Ca$_{1/2}$MnO$_3$ \cite{tomioka}, 
La$_{1/2}$Sr$_{3/2}$MnO$_4$ \cite{sternlieb,murakami}
and La$_{1/3}$Ca$_{2/3}$MnO$_3$ \cite{diaz}, 
below a certain temperatures $T_{CO}$, electronic carriers become 
localized onto specific sites, which display long-range order 
throughout the crystal structure (charge ordering). Meanwhile the 
filled Mn$^{3+}$ $e_g$ orbitals also develop long-range order 
(orbital ordering). When the temperature is further decreased 
to a much lower temperature $T_N$, an antiferromagnetic (AF) 
magnetic ordering with a zigzag pattern sets in (Fig. \ref{view}(a)).
In some others like La$_{1/2}$Ca$_{1/2}$MnO$_3$ 
\cite{radaelli}  and Nd$_{1/2}$Sr$_{1/2}$MnO$_3$ \cite{kawano},
the system first undergoes a ferromagnetic (FM) transition at the  
Curie temperature $T_C$, then enters into the CE-type charge, orbital 
and spin  ordering (COSO) state at a lower temperature $T_{CO}=T_N$. 
Theoretically, it has been proposed that the charge and orbital ordering (COO)
in half-doped manganites has a magnetic spin origin \cite{solovyev,brink}. 
However, such a theory can not be applied to those materials with 
$T_{CO}>T_N$, where the  COO is established before the spin ordering. 
Even for those whose 
$T_{CO}=T_N$, it has been shown that in a pure electronic model, 
the on-site repulsion $U$ destabilizes the CE structure
towards the rod-type (C-type) AF state (Fig. \ref{view}(b))
in realistic parameter regime of $U$ \cite{shen1}. 
Yunoki {\it et al.} \cite{yunoki} considered the effects of lattice 
distortions (LD), and found that both the noncooperative LD (NLD) and 
cooperative LD (CLD) can lead to the CE-type COSO. 
In their work 
$U$ is neglected and the calculation was performed on a 
$4\times 4\times 2$ lattice where the size effect is prominent.
Since in the absence of $U$ the CE state 
can be obtained without LD \cite{solovyev,brink}, the role of LD 
seems not very clear there. It is desirable to clarify what 
the LD results for the CE state would be destabilized by $U$.

In this work, we investigate the role of LD and large $U$ in the COSO 
in half doped manganites. For fixed magnetic ordering, by studying the 
competition between the CE and C states, we found that 
only CLD can stabilize the CE state in the presence of 
large $U$. Furthermore, without invoking the magnetic interactions, 
the CLD alone can lead to the experimentally observed COO, including 
the charge stacking effect. The magnetic ordering is the consequence of 
such a COO. 

The interaction concerning the 
lattice freedom includes two parts: $H_{lat}=H_{ep}+H_{ela}$. 
$H_{ep}$ is the coupling between lattice distortion 
and $e_g$ electrons, given by\cite{millis,hotta2}
\begin{equation}
H_{ep}=-\frac{3\lambda}{\sqrt{6}}\sum_{i\gamma\gamma^{\prime}}
c^{\gamma\dag}_{i}(\sqrt{2}\beta Q_{1i}\hat{{\bf 1}}+
Q_{2i}\sigma_x+Q_{3i}\sigma_z)_{\gamma\gamma^{\prime}}c^{\gamma^{\prime}}_i,
\label{hep}
\end{equation}
where 
$\sigma_x, \sigma_z$ are Pauli matrices, $c_{i}^{\gamma}$ is 
electron operator  of orbital $\gamma$ ($z=d_{3z^2-r^2}, 
\bar{z}=d_{x^2-y^2}$), and $Q_{1i}=\frac{1}{\sqrt{3}}(v^{x}_{i}+v^y_{i}+v^z_i), 
Q_{2i}=\frac{1}{\sqrt{2}}(v^{y}_{i}-v^x_{i}),
Q_{3i}=\frac{1}{\sqrt{6}}(2v_i^z-v^{x}_{i}-v^y_{i})$ 
are the breathing ($Q_{1i}$) and Jahn-Teller ($Q_{2i}$,
$Q_{3i}$) modes of the LD.  
Here $v^{\alpha}_i=u^{\alpha}_{i+\hat{\alpha}}-u^{\alpha}_{i-\hat{\alpha}}$,
with $u^{\alpha}_{i\pm\hat{\alpha}}$ being the $\alpha$-component of the 
displacements from the equilibrium position 
of the neighboring oxygen ion in the $\pm\alpha$ direction. The 
parameter $\beta$ is expected close to 1/2 \cite{hotta2}.
The index of spin has been omitted, which is always parallel to the 
local spin due to the strong Hund's coupling. 
Throughout the paper, we use $\alpha=x, y$ or $z$ to denote 
either the direction or the orbital state, in the latter case it 
refers to the orbital $d_{3\alpha^2-r^2}$,
whose orthogonal state is denoted as $\bar{\alpha}$. 
There are relationships $c_i^{x,y}=c_i^z/2\mp \sqrt{3}c_i^{\bar{z}}/2$ 
and $c_i^{\bar{x},\bar{y}}=\pm\sqrt{3}c_i^z/2+c_i^{\bar{z}}/2$.
$H_{ela}$ is the elastic energy and depends on the relative 
displacements of neighboring atoms with respect to the ideal 
perovskite lattice. In a unit cell of the perovskite   
A$_{1-x}$A$^{\prime}_x$MnO$_3$, there are three kinds of 
atoms: Mn, O and 
Z(=A or A$^{\prime}$). The main contribution to $H_{ela}$ 
may include the elastic energies of the neighboring Mn-O, O-Z and 
Mn-Z atoms. Up to now no studies concerning  the elastic energy of Z 
atoms has been made, however, this energy should be important, as 
without it the Z atoms can have arbitrary displacement instead of 
sitting in the center of the cubic cell cornered by eight Mn ions,
and experimental observations indicate that the Z atoms also participate 
in the LD \cite{radaelli}. 
In harmonic approximation, 
the elastic energy of the CLD may be written as
\begin{eqnarray}
H_{ela}&=&\frac{K_1}{2}\sum_{i,\kappa}
[(\boldmath{\mbox{$\delta$}}_{i}-{\bf u}_{i,\kappa})\cdot 
{\bf e}_{\kappa}]^2
+\frac{K_2}{2}\sum_{i,\xi}[(\boldmath{\mbox{$\Delta$}}_i-{\bf u}_{i,\xi})\cdot{\bf 
e}_{\xi}]^2 \nonumber \\
&\ &+\frac{K_3}{2}\sum_{i,\eta}[(\boldmath{\mbox{$\Delta$}}_{i}-\boldmath{\mbox{$\delta$}}_{i,\eta})
\cdot{\bf e}_{\eta}]^2, \label{cld}
\end{eqnarray}
where $\boldmath{\mbox{$\delta$}}, 
{\bf u}$ and $\boldmath{\mbox{$\Delta$}}$ are the 
dimensionless displacements of the Mn, O and Z ions with reference 
to the ideal perovskite lattice, 
${\bf e}_{\kappa}$, ${\bf e}_{\xi}$ and ${\bf e}_{\eta}$ are 
unit vectors along the directions of neighboring Mn-O, Z-O and
Z-Mn, respectively, with $\kappa, \xi$ and $\eta$ the indices of
neighbors. In principle, the spring constants between Z-O and Z-Mn 
depend on whether Z=A or A$^{\prime}$, here to simplify our study 
we replace them by the averaged $K_2$ and $K_3$. 
Since the distance between these neighboring atoms
are $L_{\mbox{{\scriptsize Mn-O}}}<L_{\mbox{{\scriptsize O-Z}}}<L_{\mbox{{\scriptsize Z-Mn}}}$, one expects that the spring constants $K_1>K_2>K_3$.

In the classical treatment of the LD \cite{yunoki,millis}, 
the displacements of various sites are determined by minimizing 
the total energy of the
system, $\partial H_{lat}/\partial w_i^{\alpha}=0$, where $w_i^{\alpha}$ ($w=\delta, \Delta$ or $u$) is the $\alpha$-component of 
the displacement. For harmonic $H_{ela}$, such a calculation can be
easily performed in the momentum space to get the optimized 
values of the displacements. Then after substituting these 
displacements into Eqs.(\ref{hep}) and (\ref{cld}), $H_{lat}$ reduces to 
\begin{equation}
H_{lat}^{eff}=-\epsilon_l\sum_{{\bf q},\alpha\alpha^{\prime}}f_{{\bf q}}^{\alpha\dag}G_{\alpha\alpha^{\prime}}({\bf q})f_{{\bf q}}^{\alpha^{\prime}},
\label{hlateff}
\end{equation}
where $\epsilon_l=\lambda^2/K_1$, $f_{{\bf q}}^{\alpha}$ is the 
Fourier transform of $f_i^{\alpha}=\beta n_i+m_{i}^{\alpha}$, 
with $n_i=n_i^{\alpha}+n_i^{\bar{\alpha}}$ and $m_i^{\alpha}=
n_i^{\alpha}- n_i^{\bar{\alpha}}$, and $n_i^{\alpha(\bar{\alpha})}=
c_i^{\alpha(\bar{\alpha})\dag}c_i^{\alpha(\bar{\alpha})}$.
The tensor $G=P+DR^{-1}W$, where $P, D, R$ and $W$ 
are $3\times 3$ matrices, with $P_{\alpha\alpha^{\prime}}
=\delta_{\alpha\alpha^{\prime}}
(3K_1+4K_3)S_{\alpha}^2/(3K_1S_{\alpha}^2+4K_3)$, 
$D_{xx}=h_xC,D_{xy}=-S_{xy}h_xC_z, W_{xx}=3K_1h_xC, 
W_{xy}=-3K_1S_xS_y^2C_{yz}, 
R_{xx}=4K_3(1-C^2H_x-C_z^2S_{xy}^2H_y-C_y^2S_{xz}^2H_z)+3K_2(2-C_{xy}^2-S_{xy}^2-C_{xz}^2-S_{xz}^2), R_{xy}=C_{xy}S_{xy}[4K_3(C_z^2H_x+C_z^2H_y-S_z^2H_z)+6K_2]$, where $S_{\alpha}=\sin(q_{\alpha}/2)$ and $C_{\alpha}=\cos(q_{\alpha}/2)$, $S_{\alpha\alpha^{\prime}}=S_{\alpha}S_{\alpha^{\prime}}$ and
$C_{\alpha\alpha^{\prime}}=C_{\alpha}C_{\alpha^{\prime}}$, $C=C_xC_yC_z$, 
$H_{\alpha}=4K_3/(3K_1S_{\alpha}^2+4K_3)$ and $h_{\alpha}=H_{\alpha}S_{\alpha}C_{\alpha}$. The other elements of $D, R$ and $W$ can be obtained by exchanging the indices, e.g., $D_{yy}=h_yC$ and $D_{yx}=-S_{xy}h_yC_z$, etc. 
The displacements are connected to $G$ through 
$u_{{\bf q},\alpha}^{\alpha}=-i
\lambda\sum_{\alpha^{\prime}}G_{\alpha\alpha^{\prime}}({\bf q})\langle f_{{\bf q}}^{\alpha^{\prime}}\rangle/S_{\alpha}$ and
$\delta_{{\bf q}}^{\alpha}=u_{{\bf q},\alpha}^{\alpha}-i\lambda\tan(q_{\alpha}/2)\langle f_{{\bf q}}^{\alpha}\rangle /K_1$, with 
${\bf u}_{{\bf q},\alpha}$ being the Fourier transform of 
${\bf u}_{i+\hat{\alpha}}$, and
$\boldmath{\mbox{$\Delta$}}_{{\bf q}}$ is a function of ${\bf u}_{{\bf q},\alpha}$ and 
$\boldmath{\mbox{$\delta$}}_{{\bf q}}$. 
It should be pointed out that the form of Eq.(\ref{hlateff}) 
is actually general for $H_{lat}$ with any harmonic $H_{ela}$, 
and different choice of $H_{ela}$ leads to different $G$.
For example,  the $H_{ela}$ in Ref.\ \ref{millis} includes $K_1$ and 
$K_1^{\prime}$ (the spring constant between neighboring Mn sites) 
terms, where the $G$ tensor is
$G^{(1)}_{\alpha\alpha^{\prime}}({\bf q})=\delta_{\alpha\alpha^{\prime}}(K_1+2K_1^{\prime}S^2_{\alpha})/(K_1+2K_1^{\prime})$. 
While in Refs.\ \ref{yunoki} and \ref{hotta2}, NLD and CLD yield 
$G^{(2)}_{\alpha\alpha^{\prime}}=\delta_{\alpha\alpha^{\prime}}$
and $G^{(3)}_{\alpha\alpha^{\prime}}({\bf q})=S_{\alpha}^2
\delta_{\alpha\alpha^{\prime}}$, respectively. 
Here the difference between CLD and NLD is whether
the $G$ tensor depends on ${\bf q}$ or not.

\begin{figure}[h]
\hspace{2cm}
\centerline{\epsfxsize=3.8in \epsfbox{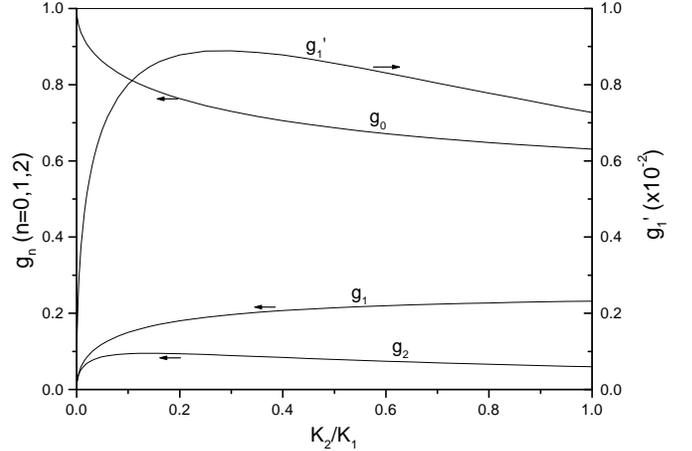}}
\caption[*]{                                                  
Calculated $g_0, g_1, g_2$ and $g_1^{\prime}$ as a function 
of $K_2/K_1$, with fixed $K_3/K_2=0.5$. 
}
\label{g0}
\end{figure}

Eq.(\ref{hlateff}) indicates that LD results in an effective 
electronic interaction. In real space, it is
\begin{eqnarray}
H_{lat}^{eff}=&-&\epsilon_l \{ \sum_{i,\alpha}[
g_0f_i^{\alpha}f_i^{\alpha}-g_1f_i^{\alpha}f_{i+\hat{\alpha}}^{\alpha}
-g_2f_i^{\alpha}f_{i+2\hat{\alpha}}^{\alpha}\nonumber \\
&+&g_1^{\prime}\sum_{\alpha^{\prime}(\neq\alpha)}f_i^{\alpha^{\prime}}
f_{i+\hat{\alpha}}^{\alpha^{\prime}}]
+{\sum_{ij,\alpha\alpha^{\prime}}}^{\prime}
G_{\alpha\alpha^{\prime}}^{ij}f_i^{\alpha}f_j^{\alpha^{\prime}}
\},\label{realspace}
\end{eqnarray}
where the sum $\sum'$ includes all the other terms. 
$g_1^{\prime}$ is found to be smaller than the coupling coefficients 
of the first several  $f_i^{\alpha}f_{i+n\hat{\alpha}}^{\alpha}$ terms, 
but larger than any other coefficients $G_{\alpha\alpha^{\prime}}^{ij}$.
Fig. \ref{g0} shows the calculated values of $g_0, g_1, g_2$ and 
$g_1^{\prime}$. From Eq.(\ref{realspace}), the main effects of the 
CLD corresponds to an effective short-range orbital-dependent coupling 
between occupied Mn sites. If $G$ is replaced by $G^{(i)}$ (i=1,2,3), then in 
the CLD cases of $G^{(1)}$ and $G^{(3)}$, there are only the $g_0$ and 
$g_1$ terms, while in the NLD case of $G^{(2)}$, there is only the $g_0$ term. 

Now let us investigate the effect of LD to the COSO at half doping.
First we see the case with fixed CE-type spin ordering. In the 
one-dimensional zigzag FM chain (see Fig. \ref{view}(a)), 
the double-exchange (DE) Hamiltonian reduces to
\begin{eqnarray}
H_{DE}&=&\sum_{i=\mbox{even} }[d_{Bi}^{\dag}(t_1c_{C,i+1}+
t_2c_{C,i-1})+\mbox{H.c.}]\nonumber \\
&\ &+U(n_{Bi}n_{Bi}^{\prime}+n_{C,i+1}^zn_{C,i+1}^{\bar{z}}),
\label{h1dde}
\end{eqnarray} 
where $d_{Bi}=c_i^x$ and $c_i^y$ for $i=4j$ and $4j+2$ (j is an integer), 
$c_{Ci}=(c_i^z,\pm c_i^{\bar{z}})^T$ for i=$4j+1$ and $4j+3$,
and $i=$even  and odd corresponds to the bridge (Mn$^{3+}$) and 
corner (Mn$^{4+}$) sites, respectively, $t_{1,2}=-(t/2,\pm\sqrt{3}t/2)$,
$n_{Bi}=d_{Bi}^{\dag}d_{Bi}$ and $n_{Bi}^{\prime}=d_{Bi}^{\prime\dag}d_{Bi}^{\prime}$, with $d_{Bi}^{\prime}$ being the orthogonal state of orbital $d_{Bi}$, 
and $n_{Ci}^{\gamma}=c_{Ci}^{\gamma\dag}c_{Ci}^{\gamma}$.
In realistic manganites, the parameter regime of the on-site repulsion
$U=10\sim 20t$, so that the system is strong correlated. 
In this regime double occupancy of electrons at a site is almost 
forbidden, and it is appropriate to use the Gutzwiller projection (GP) 
method, valid for $U\rightarrow \infty$ \cite{brink}, to take into 
account such a strong correlation effect. In GP we introduce 
constrained electrons at each site. Each electron operator 
in Eq.(\ref{h1dde}) is replaced by the corresponding projected 
operator to eliminate the double occupancy, 
$d_{Bi}\rightarrow (1-n_{Bi}^{\prime})d_{Bi}$ and 
$c_{Ci}^{\gamma}\rightarrow (1-n_i^{\bar{\gamma}})c_{Ci}^{\gamma}$ 
($\bar{\gamma}=\bar{z} (z)$ when
$\gamma=z (\bar{z})$). 
Then we use a mean-field approximation by decoupling the high order 
terms such as $c_{Bi}^{x\dag}c_{C,i+1}^zn_{i+1}^{\bar{z}}\rightarrow 
c_{Bi}^{x\dag}c_{C,i+1}^z \langle n_{i+1}^{\bar{z}}\rangle
+\langle c_{Bi}^{x\dag}c_{C,i+1}^z\rangle n_{i+1}^{\bar{z}}$.
After a Fourier transform, we have
\begin{eqnarray}
H_{DE}^{MF}&=&\sum_{k\gamma}
(d_{Bk}^{\dag}\tilde{t}_k^{\gamma}c_{Ck}^{\gamma}+
c_{Ck}^{\gamma\dag}\tilde{t}_k^{\gamma\ast}d_{Bk}
+\epsilon_{\gamma}c_{Ck}^{\gamma\dag}c_{Ck}^{\gamma})\nonumber \\
&\ &+\sum_k \epsilon^{\prime}d_{Bk}^{\prime\dag}d_{Bk}^{\prime}+E^0_{DE},
\end{eqnarray}
where 
$\tilde{t}_k^{\gamma}=\langle 1-n_{B}^{\prime}\rangle 
\langle 1-n_{C}^{\bar{\gamma}}\rangle t_k^{\gamma}$ 
with $t_k^z=-t\cos k$ and $t_k^{\bar{z}}=t\sqrt{3}i\sin k$, 
$\epsilon_{\gamma}
=-2\langle 1-n_B^{\prime}\rangle \mbox{Re}\sum_{k}\langle 
d_{Bk}^{\dag}t_k^{\bar{\gamma}}c_{Ck}^{\bar{\gamma}}\rangle$,
$\epsilon^{\prime}=
-2\mbox{Re}\sum_{k\gamma}\langle 1-n_C^{\bar{\gamma}}\rangle \langle
d_{Bk}^{\dag}t_k^{\gamma}c_{Ck}^{\gamma}\rangle$, and 
$E^0_{DE}$ is the MF energy constant.

For $H_{lat}$ of Eq.(\ref{hlateff}), in the case of CE type spin 
ordering the sum over ${\bf q}$ includes ${\bf q}=
(0,0,0),\pm(\pi/2,\pi/2,0)$ and $(\pi,\pi,0)$, and can be
denoted as $q=0,\pm\pi/2$ and $\pi$ in the 1D FM chain.
By decoupling the quartic term
$f_q^{\alpha\dag}f_q^{\alpha^{\prime}}\rightarrow 
\langle f_q^{\alpha\dag}\rangle f_q^{\alpha^{\prime}}
+f_q^{\alpha\dag}\langle f_q^{\alpha^{\prime}}\rangle$, 
Eq.(\ref{hlateff}) reduces to
\begin{equation}
H_{lat}^{MF}=\sum_{k,q}(c_{k-q}^{\dag}D_qc_{k}+\mbox{H.c.})+E^0_{lat},
\end{equation} 
where 
$c_k$ is the Fourier transform of $c_i=(c_i^z,c_i^{\bar{z}})^T$ and 
$D_q=-\epsilon_l\sum_{\alpha\alpha^{\prime}}G_{\alpha\alpha^{\prime}}(q)
\langle f_q^{\alpha^{\prime}\dag}\rangle F^{\alpha}$, 
with 
$F^{\alpha}=\beta
+\cos\phi_{\alpha}\sigma_z+\sin\phi_{\alpha}\sigma_x$ and $\langle f_q^{\alpha\dag}\rangle=\sum_{k}\langle c_{k+q}^{\dag}F^{\alpha}c_k\rangle$. 

\begin{figure}[h]
\hspace{2cm}
\centerline{\epsfxsize=3.8in \epsfbox{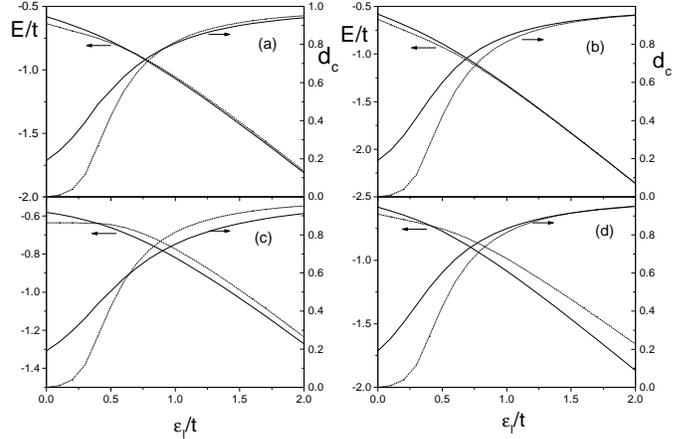}}
\caption[*]{                                                  
Energy per site in CE (solid line) and C (dotted line) 
states  as a function of $\epsilon_l/t$ and the corresponding charge 
disproportionation with (a) $G^{(1)}$, (b) $G^{(2)}$, (c) 
$G^{(3)}$ and (d) $G$, with $\beta=0.5$. 
In $G^{(1)}$ $K_1^{\prime}/K_1=0.5$, and in $G$ $K_2=2K_3=0.4K_1$.} 
\label{quantum}
\end{figure}

The full Hamiltonian $H=H_{DE}^{MF}+H_{lat}^{MF}$ can be solved by iteration. 
In the next we will make a comparison of the magnetic CE
and C states. These two states have the same magnetic energy 
$-J_{AF}$ per site, where $J_{AF}$ is
the AF magnetic superexchange between neighboring local spins, so that
the relative stability of them is independent of the parameter $J_{AF}$. 
While other magnetic states, say, the FM and layered (A)-type AF states, 
have different magnetic energies so that their
stabilities depend on $J_{AF}$, and become less stable with respect to
the CE and C state when $J_{AF}$ increases. 
In the C state there is only one effective orbital 
in each site so that $U$ has no effect 
and $H_{DE}=-t\sum_k c_k^{x\dag}c_k^{x}$ in the x-orientated FM chain. 
Such a property make the competition between the C and CE states alone
very interesting. In the absence of the electron-lattice interaction,
without $U$ the energy per site is $E^{CE}=-0.693t$ and 
$E^{C}=-0.637t$, 
when $U$ increases $E^C$ keeps unchanged while $E^{CE}$ increases 
and becomes higher than $E^C$ at about $U=5t$ \cite{shen1},  
indicating that the strong electronic 
correlation would destabilize the CE phase towards the C state.  
When the electron-lattice interaction is taken into account,
Fig. \ref{quantum} shows the energy per site and the charge disproportionation
$d_c=\langle n_{2j}-n_{2j+1}\rangle$ as a function of $\epsilon_l/t$ in the 
CE and C states. 
Fig. \ref{quantum} (a-d) corresponds to the tensor $G^{(1)}, G^{(2)}$, 
$G^{(3)}$ and the present $G$, respectively.
At $\epsilon_l=0$ the energy $E^{CE}>E^C$, the CE state is unstable.  
When the electron-lattice interaction increases,
it is found that in the NLD case of $G^{(2)}$, the CE phase always 
has higher energy than C, while in all the CLD cases of $G^{(1)}$, 
$G^{(3)}$ and $G$, there is a crossover from C to CE state with the 
increasing $\epsilon_l$. So here NLD is not enough to stabilize the
observed CE state, to obtain the CE state
the cooperative nature of the LD must be taken into account. 
The different results in the NLD and CLD cases can be understood 
from Eq.(\ref{realspace}). In the CLD cases, when a pair of neighboring 
sites are both occupied, the additional $g_1$ coupling  
favors different orbitals on the two sites. 
Since the orbitals on neighboring sites are the same in the C state 
but are different in the CE state, the latter is more favored. 
At large charge dispropornation, one of the neighboring sites is 
almost empty and the $g_1$ coupling makes no difference between C 
and CE states,
so in Fig. \ref{quantum} (a) and (c) 
the energy difference $\delta E$ of the two states no longer increases  
with further increasing $\epsilon_l$. On the other hand, $\delta E$ 
keeps increasing in (d), which is related to the $g_2$ coupling in 
Eq.(\ref{realspace}), and its effect will be discussed later.
In the calculation with $G$, it is also found that 
with different values of $K_2$ and $K_3>0$ we get qualitatively 
the same results. The calculated relative displacements of the Z, O 
and Mn sites $|\boldmath{\mbox{$\Delta$}}_{{\bf 
q}}|/|\boldmath{\mbox{$\delta$}}_{{\bf q}}|$ and 
$|\boldmath{\mbox{$\delta$}}_{{\bf q}}|/$$|u_{{\bf q},x}^x|$ 
is independent of $\epsilon_l$, the former actually depends only 
on $K_3/K_2$, and the latter decreases with increasing $K_2/K_1$ or
decreasing $K_3/K_1$. In Ref.\ \ref{radaelli} 
the ratios $|\boldmath{\mbox{$\Delta$}}_{\bf q}|/|\boldmath{\mbox{$\delta$}}_{\bf 
q}$$|\approx 0.56$ and 
$|\boldmath{\mbox{$\delta$}}_{{\bf q}}|$$/|u_{{\bf q},x}^x|$$\approx 0.93$ 
were measured
at ${\bf q}=(\pi/2,\pi/2,0)$. 
In the present calculation with 
$K_2=2K_3=0.4K_1$ 
the two ratios are 0.57 and 0.90, quite close to that in 
Ref.\ \ref{radaelli}.

\begin{figure}[h]
\hspace{2cm}
\centerline{\epsfxsize=3.8in \epsfbox{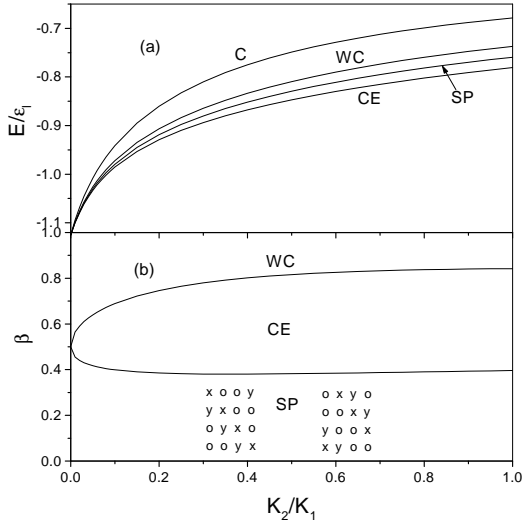}}
\caption[*]{                                                  
In the classical treatment of electrons and with 
fixed $K_3/K_2=0.5$, (a) engery per site in the CE, WC, C and SP 
states at $\beta=0.5$ as a function of $K_2/K_1$, (b) 
phase diagram from MC 
simulation. The $4\times 4\times 2$ unit cell of SP is shown in (b), 
where the two $4\times 4$ lattices are in successive x-y planes,
and x, y, o represent orbitals $d_{3x^2-r^2}$, $d_{3y^2-r^2}$ and 
a hole.
}
\label{classical}
\end{figure}

At strong electron-phonon interaction ($\epsilon_l\gg t$), the charge 
disproportionation tends to 1, the electrons become localized
and can be treated as classical objects.
A classical treatment of electrons can simplify the study 
considerably and show clearly the effect of CLD,  
and in fact should be appropriate in some manganites in which 
the charge difference between neighboring Mn sites is close 
to 1 \cite{murakami}. In the classical case, $n_{\bf q}=
(\delta_{{\bf q},0}+\delta_{{\bf q},{\bf Q}_{\|}})\sqrt{N}/2$ 
in both the C and CE states, where ${\bf Q}_{\|}=(\pi,\pi,0)$ 
and $N$ is the total number of Mn sites, and $m_{\bf q}^x=n_{\bf q}$, 
$m_{\bf q}^y=m_{\bf q}^z=-n_{\bf q}/2$ in the C state,
$m_{\bf q}^{x,y}=n_{\bf q}/4\pm 3(\delta_{{\bf q},{\bf Q}_{\|}/2}+
\delta_{{\bf q},-{\bf Q}_{\|}/2})\sqrt{N}/8, m_{\bf q}^z=-n_{\bf q}/2$ 
in the CE state. For $G^{(i)} (i=1,2,3)$,
the energies obtained from $f_{\bf q}^{\alpha}=\beta n_{\bf q}+
m_{\bf q}^{\alpha}$ and Eq.(\ref{hlateff}) are the same in C and CE states. 
If we further take the Wigner crystal (WC) state into account, which has 
the same COSO as that of CE in the xy plane,
but along the z direction the charge density is altering instead of 
stacking, and $f_{\bf q}^{\alpha,\mbox{{\scriptsize WC}}}=[f_{\bf 
q}^{\alpha,\mbox{{\scriptsize CE}}}(1+e^{iq_x})
+f_{{\bf q}-{\bf Q}_z}^{\alpha,\mbox{{\scriptsize CE}}}(1-e^{iq_x})]/2$ with 
${\bf Q}_z=(0,0,\pi)$, then the energy difference between   
CE (or C) and WC states are 
$\epsilon_l(\beta-1/2)^2K_1^{\prime}/(2K_1+4K_1^{\prime})-J_{AF}$, $-J_{AF}$ 
and $\epsilon_l(\beta -1/2)^2/4-J_{AF}$ for $G^{(1)}$, $G^{(2)}$ and $G^{(3)}$.
So that in the cases of $G^{(i)}$ ($i=1,2,3$),
without $J_{AF}$ WC should be more stable than CE state, 
and for $\beta=1/2$ a finite $J_{AF}$ would yield stable 
degenerate CE- and C-
stacking states.
On the other hand,  Fig. \ref{classical}(a) shows the 
energy of the C, CE, WC and stripe phase (SP) states with $G$ at 
$\beta=1/2$ and $J_{AF}=0$, in which CE state has the lowest energy. 
A Monte Carlo (MC) simulation on $8\times 8\times 8$ lattice 
with periodic boundary conditions, in which 
we consider three possible electronic states on a Mn site including
occupied by elongated orbitals $d_{3x^2-r^2}$, $d_{3y^2-r^2}$ and 
unoccupied, is performed in real space to find the charge and orbital 
configuration with the lowest energy. The compressed orbitals such 
as $d_{x^2-y^2}$ are not taken into account due to the anharmonic 
effects \cite{millis,khomskii}. For a given configuration,  
we calculate its $f_{\bf q}^{\alpha}$ in momentum space and get 
the energy through Eq.(\ref{hlateff}). Fig. \ref{classical}(b) 
is the calculated phase diagram. Note that phase diagram is
obtained not by comparing the several states in Fig. \ref{classical}(a), 
instead, each state here has the lowest
energy among all the possible charge and orbital configurations 
within the range of consideration. For $\beta$ close to $1/2$,  
the obtained 
COO is the same as that under CE-type spin environment 
shown in Fig. \ref{view}(a). 
The striking feature of 
charge stacking (CS) along the z direction is 
reproduced. Such a stacking is usually attributed 
to $J_{AF}$ \cite{yunoki}, yet here our calculated 
results provides another possible explanation, that 
the CLD may also lead to the CS. It is worth mentioning that the 
above simulation can be easily generalized to $2/3$ doping, 
where the obtained COO for $\beta$ close to $1/2$ is a CS state same as 
that observed in  Ref.\ \ref{diaz}, and such a CS can not be explained 
by $J_{AF}$.  The COO at half-doping may be interpreted by the effective 
interaction shown in Eq.(\ref{realspace}). 
The COO in the xy plane  can be explained by the $g_1$ and $g_2$ terms. 
The $g_1$ term favors the charge ordering peaked at $(\pi,\pi)$. In such 
an ordering, if site $i$ is occupied, then $i+\hat{\alpha}$ ($\alpha=x$ 
or $y$) is empty and $i+2\hat{\alpha}$ is occupied. 
The $g_2$ coupling 
between the occupied sites $i$ and $i+2\hat{\alpha}$ then prefers the 
orbitals in the two sites to be different, thus the desired 
in-plane COO is formed. The stacking in the 
z-direction is related to the $g_1^{\prime}$ coupling which 
favors neighboring sites occupied by the same orbitals. When $\beta$ 
is close to $1/2$, along the z direction the $g_1$ and $g_2$ coupling is 
effectively very weak as at an occupied site $f_i^z=\beta-1/2$ is small, 
then the $g_1^{\prime}$ term is dominant. Note that here the COO is 
obtained without invoking magnetic interactions, so that the COO 
transition temperature $T_{CO}$ can be higher than the magnetic 
transition temperature $T_N$, in agreement with the experiments 
in some doped manganites \cite{tomioka,sternlieb,murakami,diaz}. 
Since the effect discussed above exists beyond the classical limit, 
in more general cases the CLD should also favor such a COO. 
Once this COO is built, the CE-type zigzag magnetic ordering below 
$T_N$ can be understood from the competition between the anisotropic 
electronic hopping and $J_{AF}$.  For an occupied 
site $i$ of orbital $d_{3x^2-r^2}$ ($d_{3y^2-r^2}$), 
the electronic hopping between sites $i$ and $i+\hat{x}$ ($i+\hat{y}$) 
leads the spins of these two sites to be parallel. On the other hand,
the spins of $i$ and its neighbors in the other two directions are 
antiparallel as the electronic hopping integrals in these two directions 
are much smaller and not enough to overcome $J_{AF}$.  
In this way naturally the CE-type zigzag magnetic ordering 
shown in Fig. \ref{view} (a) is obtained. 
In this picture of the COSO,  appropriate at least for 
those whose $T_{CO}>T_N$, 
COO has its origin of the cooperative nature of the 
lattice distortion, and the spin ordering is the consequence of such a COO.

This work is supported by a grant from Texas ARP 
grant (ARP-003652-0241-1999), the Robert A. Welch 
Foundation and the Texas
Center for Superconductivity at the University of Houston.

\end{document}